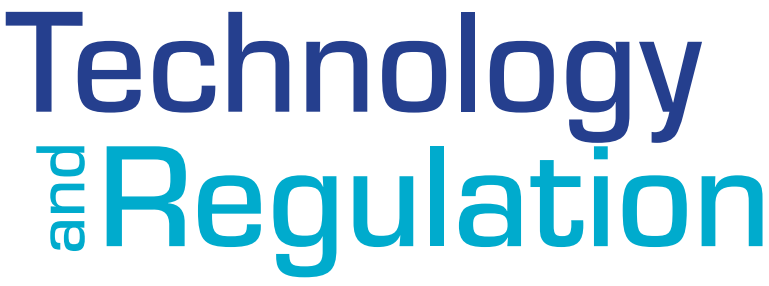

# Digital Discrimination in Dating Apps and the Dutch Breeze case

| | |
|---|---|
| **Author(s)** | Tim de Jonge & Frederik J. Zuiderveen Borgesius |
| **Contact** | tim.dejonge@ru.nl |
| **Affiliation(s)** | Tim de Jonge is a PhD Candidate Fairness in Machine Learning at the iHub, Radboud University<br>Frederik J. Zuiderveen Borgesius is Professor if ICT and law at the iHub, Radboud University |

## Abstract

**In 2023, the Netherlands Institute for Human Rights, the Dutch non-discrimination authority, decided that Breeze, a Dutch dating app, was justified in suspecting that their algorithm discriminated against dark-skinned users. Consequently, the Institute decided that Breeze must prevent this discrimination based on ethnicity. This paper analyses the decision and explores three questions.What are the main points of the Breeze decision? Is the discrimination based on ethnicity in Breeze's matching algorithm illegal? We also explore a more general question: how can dating apps mitigate or stop discrimination in their matching algorithms? We illustrate the legal and technical difficulties dating apps face in tackling discrimination and highlight some promising solutions. We analyse the Breeze decision in-depth, combining insights from computer science and law. We discuss the implications of this judgment for scholarship and practice in the field of fair and non-discriminatory machine learning.**

## 1. Introduction

In 2023, the Netherlands Institute for Human Rights ('the Institute') decided that Breeze, a Dutch dating app, was justified in suspecting that their algorithm discriminated against dark-skinned users.[1] Moreover, the Institute decided that Breeze must prevent this discrimination. The Institute also decided that it does not count as positive discrimination if Breeze de-biases its digital system.

In this paper, we explore three questions. (i) What are the main points of the Breeze decision? (ii) Is the discrimination on the basis of ethnicity in Breeze's matching algorithm illegal? (iii) How can dating apps mitigate or stop discrimination in their matching algorithms?

The paper makes the following contributions to the literature. First, this is the first in-depth scholarly discussion of the Breeze decision, as far as we know.[2] Second, we combine insights from law and computer science. In addition to our legal analysis, we investigate, from a technical perspective, the practical usability of the metric suggested by the Institute, and the legal applicability of metrics provided by the Fair Machine Learning community. We illustrate the legal and technical difficulties dating apps face in tackling discrimination and highlight promising approaches.

The Breeze decision is important, because this is one of the first legal decisions that applies traditional European non-discrimination law to digital discrimination (see Section 3.2). The Breeze decision shows, as scholars have argued already,[3] that non-discrimination law can be applied to digital or AI-driven discrimination.

This paper can be relevant for legal scholars who are interested in digital discrimination and the quantitative considerations regarding discrimination, and for people working with digital systems with discrimination risks.

A note on terminology. The Institute speaks of 'users with a dark skin colour or from non-Dutch origin' and 'users with a light skin colour or from non-Dutch origin'.[4] For ease of reading, we speak of 'dark-skinned users' and 'light-skinned users' in the paper.

The paper is structured as follows. In Section 2, we summarize the Institute's decision. In Section 3, we provide a legal analysis of the decision. We also discuss the question whether Breeze discriminates illegally. In Section 4 and 5, we investigate how discrimination could be measured and mitigated in dating apps. Section 6 concludes.

## 2. Overview of the Breeze Case

In this section, we summarize the case as it is presented by the Institute. Before we investigate the decision, let us take a closer look at the two main parties in this story.

---

1. College voor de Rechten van de Mens, 'Dating-App Breeze Mag (En Moet) Algoritme Aanpassen Om Discriminatie Te Voorkomen' (6 September 2023) https://www.mensenrechten.nl/actueel/nieuws/2023/09/06/dating-app-breeze-mag-en-moet-algoritme-aanpassen-om-discriminatie-te-voorkomen accessed 13 September 2024; College voor de Rechten van de Mens, Oordeelnummer 2023-82 (6 September 2023) https://oordelen.mensenrechten.nl/oordeel/2023-82 accessed 13 September 2024.
2. Weerts et al. discuss some aspects of the decision; Hilde Weerts and others, 'The Neutrality Fallacy: When Algorithmic Fairness Interventions Are (Not) Positive Action' in Proceedings of the 2024 ACM Conference on Fairness, Accountability, and Transparency (Association for Computing Machinery 2024) https://dl.acm.org/doi/10.1145/3630106.3659025 accessed 3 April 2025.
3. *See* e.g. Jeremias Adams-Prassl, Reuben Binns, and Anne Marie Kelly-Lyth, 'Directly Discriminatory Algorithms' (2023) 86(1) *The Modern Law Review* 144; Frederik J. Zuiderveen Borgesius, 'Price Discrimination, Algorithmic Decision-Making, and European Non-Discrimination Law' (2020) 31(3) *European Business Law Review* 401-222.
4. College voor de Rechten van de Mens, 'Breeze Social B.V. Discrimineert Niet, Als Zij Maatregelen Neemt Die Voorkomen Dat Haar Algoritme Gebruikers Met Een Niet-Nederlandse Afkomst of Donkere Huidskleur Benadeelt' Oordeelnummer 2023-82, e.g. para 5.3 https://oordelen.mensenrechten.nl/oordeel/2023-82 accessed 13 September 2024; *All translations of cited phrases from the decision are by the authors.*

The Netherlands Institute for Human Rights is an independent Dutch institute, funded by the state. The Institute is tasked with protecting human rights, increasing awareness of human rights, and promoting the safeguarding of human rights.[5] Among the tasks of the Institute is reviewing cases of possible discrimination. An individual can submit a complaint about discrimination, and a company can ask the Institute to assess whether a certain practice constitutes illegal discrimination or not.[6] A decision of the Institute (as discussed in this paper) resembles a court judgment, but is advisory, rather than legally binding.

Breeze is a Dutch dating app, founded in 2020 by graduates from the Technical University Delft.[7] Breeze has roughly 30.000 active users per month.[8] On Breeze, users are shown profiles made by other users, containing photos, age, and whatever information the user volunteers. A user can express their interest in a date by *liking* the other user. If both users like each other, they *match*. On Breeze, once users match, they are prompted to set up a date. Breeze has partnered with various bars, restaurants, and similar venues suitable for dates, and users can pick one of these venues, along with a time. Breeze charges users for each match and restricts users from backing out of dates.

Breeze does not offer a chat function. Most other apps first present matches with a chat window, where matches can get to know each other, and decide whether they want to date. On Breeze, liking someone means there is a realistic chance of going on a date with that person.

Like many other dating apps, Breeze employs a matching algorithm to show its users profiles of other users. This algorithm uses the information a user provides on their profile along with the preferences of previous users to show the most promising prospects to a user.[9]

Breeze only shows a user a handful of user profiles per day. Many other dating apps do not limit the number of profiles shown: the app will rank all profiles, either by a matching algorithm (e.g. Tinder[10]) or by proximity (e.g. Happn[11]), and the user is free to respond to as many profiles as the user wants.

The limited number of interactions between user and algorithm on Breeze places great importance on their matching algorithm. On other apps, the user can circumvent the algorithm to a certain extent by continually rejecting suggestions, until the user finds options that fit their taste. On Breeze, this is not an option. If Breeze's matching algorithm does not serve a particular group of users well, these users might turn to other means to find romance.

On Breeze, users cannot choose options such as 'I only want to see light-skinned people' or 'filter out people with an Asian background'.[12] Breeze does not ask its users for information about their ethnicity.[13] Breeze's algorithm cannot access users' pictures.

5. College voor de Rechten van de Mens, 'About the Institute' https://www.mensenrechten.nl/english accessed 14 September 2024; European Network of National Human Rights Institutions, 'Implementing the Council of Europe Recommendation on National Human Rights Institutions: The State of Play, The situation in Netherlands' (2023) https://ennhri.org/wp-content/uploads/2023/05/Netherlands-Country-Report-CoE-NHRI-Rec-ENNHRI-Baseline.pdf 14 September 2024.
6. Wet College voor de rechten van de mens (Statute on the Netherlands Institute for Human Rights), art 10(2)(b) http://wetten.overheid.nl/jci1.3:c:BWBR0030733 accessed 14 September 2024.
7. *See* Breeze Social, https://breeze.social accessed 14 September 2024.
8. College voor de Rechten van de Mens, 'Breeze Social B.V. Discrimineert Niet, Als Zij Maatregelen Neemt Die Voorkomen Dat Haar Algoritme Gebruikers Met Een Niet-Nederlandse Afkomst of Donkere Huidskleur Benadeelt' Oordeelnummer 2023-82, para 3.2 https://oordelen.mensenrechten.nl/oordeel/2023-82 accessed 14 September 2024.
9. College voor de Rechten van de Mens (n 4) para 3.3, 3.4.
10. *See* Tinder, https://tinder.com accessed 14 September 2025.
11. *See* Happn, https://happn.com accessed 14 September 2025.
12. Some other dating apps, for example in the US, offer such possibilities. *See* Jevan A. Hutson, Jessie G. Taft, Solon Barocas, and Karen Levy, 'Debiasing Desire: Addressing Bias & Discrimination on Intimate Platforms' (2018) 2 *Proceedings of the ACM on Human-Computer Interaction*, CSCW, Article 73, 1–18; Apryl Williams, *Not My Type: Automating Sexual Racism in Online Dating* (Stanford University Press 2024) 268.
13. College voor de Rechten van de Mens (n 4) para 3.6.

## 2.1 Breeze's Request

Breeze told the institute that it became suspicious of possible discrimination by its app in two ways. First, when Breeze surveyed their users to find ways to improve their app, some users responded that they were not seeing enough ethnic diversity in their suggested matches.[14] Second, Breeze checked its own app and discovered that the app is probably less likely to show dark-skinned users than light-skinned users. Breeze's user base is predominantly light-skinned.[15]

Breeze suspected that their algorithm might unintentionally be discriminating, causing this lack of diversity: the algorithm might show the profiles of their dark-skinned users more frequently than the profiles of their light-skinned users. Breeze chose not to keep their suspicions secret but contacted the Institute for advice on preventing the suspected discrimination in their algorithm.

Breeze alleges that people who are looking for a date often prefer people of the same ethnicity as themselves.[16] Breeze further claims that their algorithm is more likely to show profiles of people who have a higher *match percentage*. Roughly speaking, Breeze claims that popular users get more attention than unpopular users. The algorithm is self-learning and a type of black box. Breeze told the Institute that they do not know exactly how the algorithm calculates the match percentage.[17]

The Breeze user base is predominantly light-skinned.[18] From Breeze's assumption that many people are more likely to prefer dating within their own ethnicity, it follows that dark-skinned people have a lower match percentage on Breeze. Considering that the algorithm prefers to show people with high match percentage, Breeze says that their algorithm would show light-skinned users more frequently than dark-skinned users.[19]

Breeze aims for a more diverse user base, and posed the following two questions to the Institute:

Q1: Does non-discrimination law allow Breeze to adapt the functioning of the algorithm to promote diversity among Breeze users?[20]

Q2: To what extent does the law allow Breeze to not only compensate for ethnicity discrimination, but to take measures that increase the matching chances of dark-skinned users? (Such measures could be interpreted as favouring dark-skinned users over other users on the basis of their skin colour or origin)[21]

## 2.2 The Decision on Breeze

To answer the questions Breeze posed, the Institute relies on a close reading of Dutch discrimination law. First, the Institute assesses whether discrimination law applies to the presented case. The law does not prohibit all instances of discrimination. The Institute says that, from a legal perspective, app users are free to have preferences for people of a certain ethnicity when choosing a date.[22]

To support the claim that discrimination law applies in this case, the Institute invokes Article 7(1)(a) of the Equal Treatment Act, which prohibits companies from discrimination when providing access to a good

14. College voor de Rechten van de Mens (n 4) para 5.9.
15. College voor de Rechten van de Mens (n 4) para 5.9.
16. College voor de Rechten van de Mens (n 4) para 5.9.
17. College voor de Rechten van de Mens (n 4) para 3.5.
18. College voor de Rechten van de Mens (n 4) para 5.9.
19. College voor de Rechten van de Mens (n 4) para 3.4 and 5.9.
20. College voor de Rechten van de Mens (n 4) paras 4, 5.15.
21. College voor de Rechten van de Mens (n 4) paras 4, 5.16. We slightly amended the phrasing of the question for more clarity.
22. College voor de Rechten van de Mens (n 4) para 5.13. For a non-legal discussion on whether it is acceptable for people to have preferences for certain ethnicities when dating ('sexual discrimination'), *see* Apryl Williams, *Not My Type: Automating Sexual Racism in Online Dating* (Stanford University Press 2024) 268.

or service.[23] The Institute chooses a broad view of 'access': if Breeze functions sufficiently worse for dark-skinned users, this can be interpreted as lesser access for dark-skinned users.[24]

The Equal Treatment Act distinguishes direct from indirect discrimination.[25] Roughly speaking, direct discrimination takes place when an individual is treated differently on the basis of their ethnicity.[26]

Indirect discrimination occurs when a practice appears neutral at first glance but ends up discriminating against people with a certain ethnic origin (or another legally protected characteristic).[27] Indirect discrimination can sometimes be justified, namely if the alleged discriminator can rely on an objective justification. We return to the possibility of an objective justification in Section 3.1. For now, it suffices to understand that the legal treatment of direct and indirect discrimination differs, and so, that the Institute must determine whether Breeze discriminates directly or indirectly. For the law it is not relevant whether discrimination happens on purpose or by accident.

The Institute checks first whether Breeze or its algorithm engages in direct discrimination. Breeze detailed that their matching algorithm does not have access to any information indicating ethnicity. Although Breeze's users' profiles contain pictures, Breeze keeps these pictures separate from the matching algorithm. The algorithm uses the users' profile texts and the users' like-behaviours, which may result in different outcomes for different demographics. The algorithm does not – at least not directly – rely on data related to users' ethnicity or skin colour in determining the match probability. Therefore, the Institute concludes that Breeze does not discriminate directly.[28]

Next, the Institute checks whether Breeze discriminates indirectly. Under the Equal Treatment Act, indirect discrimination occurs if an apparently neutral practice particularly affects persons of a certain ethnicity compared to other persons.[29] The algorithm is apparently neutral: it does not have access to any data indicating ethnicity, and it is created to make matches between users.

The Institute accepts the assumptions that Breeze made in its request, and concludes that the algorithm shows profiles of dark-skinned users less frequently. This is a negative impact and constitutes indirect discrimination, says the Institute.[30]

The law provides Breeze room to try to justify this discriminatory outcome: if Breeze has an *objective justification* for the discrimination, it is not illegal. We outline the details of this justification in Section 3.1. In this case however, Breeze's request did not include an attempt at justification. The Institute holds that it is not relevant to Breeze's questions whether this indirect discrimination can be justified. Therefore, the Institute does not answer the question whether this discrimination would be illegal.[31]

23. Algemene Wet Gelijke Behandeling (Dutch Equal Treatment Act) http://wetten.overheid.nl/jci1.3:c:BWBR0006502 accessed 15 September 2024; International Commission of Jurists, *Netherlands: Equal Treatment Act* (2004, as amended, unofficial English translation) https://www.icj.org/wp-content/uploads/2013/05/Netherlands-Equal-Treatment-Act-2004-eng.pdf accessed 22 April 2025; From that translation: Article 7(1)(a): 'It is unlawful to discriminate in offering goods or services, in concluding, implementing or terminating agreements thereon, and in providing educational or careers guidance if such acts of discrimination are committed: a. in the course of carrying on a business or practising a profession (...)'.
24. College voor de Rechten van de Mens (n 4) para 5.7-5.10.
25. College voor de Rechten van de Mens (n 4) para 5.11.
26. For instance, if a company says it will not recruit employees with a certain skin colour, this is an example of direct discrimination. Case C-54/07 *Centrum voor gelijkheid van kansen en voor racismebestrijding v Firma Feryn NV* [2008] CJEU, ECLI:EU:C:2008:397, para 25.
27. See generally on the concept of indirect discrimination: Christa Tobler, *Indirect Discrimination: A Case Study Into the Development of the Legal Concept of Indirect Discrimination under EC Law*, vol 10 (Antwerp: Intersentia 2005); Evelyn Ellis and Philippa Watson, *EU Anti-Discrimination Law,* 148-155 (Oxford: Oxford University Press 2012).
28. College voor de Rechten van de Mens (n 4) para 5.11, 5.12.
29. Article 1(c) of the Equal Treatment Act defines indirect discrimination as follows: 'indirect discrimination: if an apparently neutral provision, criterion or practice particularly affects persons of a particular religion, belief, political affiliation, race, sex, nationality, heterosexual or homosexual orientation or marital status compared to other persons.' The possibility for an objective justification is provided in article 2(1).
30. College voor de Rechten van de Mens (n 4) para 5.11, 5.13.
31. College voor de Rechten van de Mens (n 4) para 5.14.

From here, the Institute answers the questions that Breeze asked.

Q1: Does non-discrimination law allow Breeze to adapt the functioning of the algorithm to promote diversity among Breeze users?[32]

The Institute establishes that Breeze did not provide an objective justification for the indirect discrimination in their matching algorithm. Therefore, says the Institute, Breeze must remove this discriminatory effect, or prevent this discrimination in the future. Breeze is not just *allowed* to adapt the algorithm to promote diversity, they are *required* to adapt the algorithm to prevent this discriminatory effect, says the Institute.[33]

Q2: To what extent does the law allow Breeze to not only compensate for ethnicity discrimination, but to take measures that increase the matching chances of dark-skinned users? (Such measures could be interpreted as favouring dark-skinned users over other users based on their skin colour or origin)[34]

The Institute answers that the measures Breeze would take to prevent indirect discrimination in their algorithm should not be considered positive discrimination as defined in the Equal Treatment Act.

The Institute interprets Breeze's request as asking whether Breeze is allowed to increase the match percentage of dark-skinned users to prevent a discriminatory effect. If Breeze takes measures to prevent discrimination, this does not assign a privileged position to dark-skinned users, but merely a more equal position. Therefore, the Institute says that this does not constitute positive discrimination at all. Now that we have summarised the case and the decision, we proceed with analysis of the Institute's decision.

## 3. Legal Commentary

### 3.1 Initial Assumptions

Breeze says that its app probably recommends light-skinned users more often than other users.[35] The Institute accepts this claim, but neither points to research supporting this assumption. Some research on online dating in the United States indeed suggests that people are likely to prefer dating people from their own ethnicity.[36] Potârcă and Mills reproduce this finding in Europe, and further find that dark-skinned people are more likely to voice preference for light-skinned people than for people of other ethnicities outside of their own.[37] Hence, it seems likely that light-skinned users will, on average, match more frequently than dark-skinned users on the Breeze app.

Breeze also says that their algorithm is likely to recommend users with a high match percentage more frequently. [38] With this claim, Breeze appears to allude to *popularity bias*. Popularity bias occurs when an algorithm recommends popular items even more frequently than their popularity would warrant.[39] The algorithm in a dating app is a particular case of a broader class of algorithms known as 'recommender systems'. Popularity bias is a well-known risk of recommender systems: Boratto remarks that recommender systems 'tend to suggest popular items more than niche items, even when the latter would be of interest'.[40]

---

32. College voor de Rechten van de Mens (n 4) para 4, 5.15.
33. College voor de Rechten van de Mens (n 4) para 5.16.
34. College voor de Rechten van de Mens (n 4) paras 4, 5.16; We slightly amended the phrasing of the question for more clarity.
35. College voor de Rechten van de Mens (n 4) para 5.9.
36. Wei-Chin Hwang, 'Who Are People Willing to Date? Ethnic and Gender Patterns in Online Dating' (2013) 5 *Race and Social Problems* 28–40; Günter J. Hitsch, Ali Hortaçsu and Dan Ariely, 'What Makes You Click?—Mate Preferences in Online Dating' (2010) 8 *Quantitative Marketing and Economics* 393–427; Gina Potârcă and Melinda Mills, 'Racial Preferences in Online Dating across European Countries' (2015) 31 *European Sociological Review* 326–341.
37. Potârcă and Mills (n 36).
38. College voor de Rechten van de Mens (n 4) para 3.3, 3.4.
39. Himan Abdollahpouri and Masoud Mansoury, 'Multi-Sided Exposure Bias in Recommendation' (arXiv, 2020) http://arxiv.org/abs/2006.15772 accessed 1 May 2025.
40. Ludovico Boratto, Gianni Fenu and Mirko Marras, 'Connecting User and Item Perspectives in Popularity Debiasing for Collaborative Recommendation' (2021) 58(1) *Information Processing & Management* 102387.

Popularity bias can directly reduce the usefulness of a dating app: a user of a dating app could be looking for someone that particularly fits them, rather than someone that is popular on the dating app. Previous research has established that popularity bias can lead to discrimination.[41] Given that Breeze does not indicate that they have countered popularity bias in any way, it is plausible that Breeze's algorithm does show popularity bias. A statistical test might be able to point this out, but Breeze does not detail the degree to which the algorithm prefers showing profiles with higher match percentage.[42]

### 3.2 Application of Discrimination Law

As noted in the introduction, this is one of the first decisions in Europe in which a court or a regulator applies non-discrimination law to digital or AI-driven discrimination. One earlier case was the 2020 Deliveroo case in Italy, in which a court decided that Deliveroo's algorithm had indirectly discriminatory effects for certain groups of workers.[43]

There were some other earlier cases that were about, loosely speaking, forms of digital discrimination, but in those decisions, courts did not apply non-discrimination law. For example, the CJEU applied the General Data Protection Regulation (GDPR)[44] in cases about automated decision-making and related transparency rights for affected people.[45] In the Dutch Syri[46] and Uber cases,[47] courts applied the GDPR on automated fraud detection and on automated decision-making about drivers.

In the Breeze case, the Institute invokes Article 7(1)(a) of the Equal Treatment Act, which prohibits companies from discrimination when providing access to a good or service.[48] The Institute chooses a broad view of 'service': an economic activity, for which one can receive compensation.[49] The Institute's broad view of service is in line with case law of the Court of Justice of the European Union (CJEU). The CJEU says that the scope of the non-discrimination directive on which the Dutch legal provision is based ' 'cannot be defined restrictively'.[50]

At first glance, it may not be obvious that this provision applies. Initially, Breeze is free: users only pay when they go on a date. The Institute establishes that Breeze provides an economic service: Breeze takes an active role in mediating interaction between its users. Beyond just individual users' preferences, Breeze's matching algorithm plays an important role in establishing contact between users. In sum, the Institute concludes that the Equal Treatment Act applies to Breeze's request.

41. Michael D. Ekstrand and others, 'Fairness in Recommender Systems' in Francesco Ricci, Lior Rokach and Bracha Shapira (eds), *Recommender Systems Handbook* (3rd edn, Springer US 2022) 679–707 https://doi.org/10.1007/978-1-0716-2197-4_18 accessed 22 April 2025; Yashar Deldjoo and others, 'Fairness in Recommender Systems: Research Landscape and Future Directions' (2024) 34(1) *User Modeling and User-Adapted Interaction* 59–108.
42. College voor de Rechten van de Mens (n 4) para 5.9.
43. *See* Vincenzo Pietrogiovanni, 'Deliveroo and Riders' Strikes: Discriminations in the Age of Algorithms' [2021] 7(3) *International Labor Rights Case Law* 317–321; Ilaria Purificato, 'Behind the Scenes of Deliveroo's Algorithm: The Discriminatory Effect of Frank's Blindness' [2021] 14(1) *Italian Labour Law e-Journal* 169–194.
44. Regulation (EU) 2016/679 of the European Parliament and of the Council of 27 April 2016 on the protection of natural persons with regard to the processing of personal data and on the free movement of such data, and repealing Directive 95/46/EC (General Data Protection Regulation) [2016] OJ L119/1.
45. *SCHUFA Holding AG v [Land] Hessen* (Case C-634/21) [2023] CJEU, ECLI:EU:C:2023:957, 7 December 2023; *Dun & Bradstreet Austria GmbH* (Case C-203/22) [2025] CJEU, ECLI:EU:C:2025:117, 27 February 2025; *See also* Luka Metikoš and Jef Ausloos, 'The Right to an Explanation in Practice: Insights from Case Law for the GDPR and the AI Act' [2025] 17(1) *Law, Innovation and Technology* 205–240.
46. Marvin van Bekkum and Frederik J. Zuiderveen Borgesius, 'Digital Welfare Fraud Detection and the Dutch SyRI Judgment' [2021] 23(4) *European Journal of Social Security* 323–340.
47. Raphaël Gellert, Marvin van Bekkum and Frederik Zuiderveen Borgesius, 'The Ola & Uber Judgments: For the First Time a Court Recognises a GDPR Right to an Explanation for Algorithmic Decision-Making' (EU Law Analysis, 2021) https://eulawanalysis.blogspot.com/2021/04/the-ola-uber-judgments-for-first-time.html accessed 22 April 2025.
48. Algemene Wet Gelijke Behandeling (Dutch Equal Treatment Act) http://wetten.overheid.nl/jci1.3:c:BWBR0006502 accessed 22 April 2025; International Commission of Jurists, *Netherlands: Equal Treatment Act* (2004, as amended, unofficial English translation) https://www.icj.org/wp-content/uploads/2013/05/Netherlands-Equal-Treatment-Act-2004-eng.pdf accessed 22 April 2025 (n 23).
49. College voor de Rechten van de Mens (n 4) para 5.7-5.10.
50. Case C-83/14 *CHEZ Razpredelenie Bulgaria AD v Komisia za zashtita ot diskriminatsia* ECLI:EU:C:2015:480 (CJEU, 16 July 2015), para 42.

### 3.3 Positive Discrimination

The primary lesson from the case is the Institute's decision that Breeze does not positively discriminate when they mitigate discrimination in their system. The relevant statute in this case is the Dutch Equal Treatment Act,[51] which is largely based on EU non-discrimination directives. Article 2.3 of the Act defines positive discrimination as follows:

> The prohibition on discrimination contained in this Act does not apply if the aim of the discriminatory measure is to place women or persons belonging to a particular ethnic or cultural minority group in a privileged position in order to eliminate or reduce existing inequalities connected with race or sex and the discrimination is in reasonable proportion to that aim.[52]

An important condition for positive discrimination to be allowed is that the measure 'is in reasonable proportion' to the aim of reducing existing inequality.[53]

We illustrate the Dutch rules on positive discrimination with a case regarding the Eindhoven University of Technology. In 2019 that university reserved all academic positions for women, for a limited period.[54] This measure was intended to increase gender equality among the higher academic echelons. The Institute declared this program incompatible with (positive) discrimination law. Roughly summarized, the university's program was too blunt. The university then changed its policy: it restricted the program to academic disciplines where the percentage of women was less than 35%. After the restriction, the program was approved by the Institute and is still running.[55]

Weerts et al.[56] investigate whether interventions as suggested by Breeze should be considered positive action under EU law. Adaptations to algorithms or AI systems to make them less discriminatory are often called 'fairness interventions' in computer science circles.[57] There is no specific case law on fairness interventions. Weerts et al. find that under current case law, courts would probably treat fairness interventions as quota. The CJEU has previously found that quotas are only legal if the quota serves only as tiebreaker, if the quota includes a saving clause, and if the quota is proportional to the harm they aim to prevent. [58] The CJEU initially proposed these restrictions in the context of job applications.

Weerts et al. suggest that instead, fairness interventions should not be taken as positive action when these interventions address pre-existing bias in a digital system. They further illustrate this point with several ways a model can introduce bias. Weerts et al. agree with the Institute's decision in the Breeze case. They suggest a general rule: if we have a biased system, fixing the bias does not mean positive discrimination.

### 3.4 Did Breeze discriminate illegally?

The Institute focused its decision on the questions asked by Breeze, and chose not to investigate whether Breeze's discrimination would be illegal. The Institute stated that, whether or not the discrimination is illegal, the law allows Breeze to mitigate discrimination in their matching algorithm.[59] The following section describes what the analysis could look like, if the Institute had investigated whether Breeze's discrimination was illegal. While we do not have insight into Breeze's algorithm or their procedures, we can outline the arguments that Breeze could present in such a case.

---

51. Algemene Wet Gelijke Behandeling (Dutch Equal Treatment Act) (n 23).
52. Algemene Wet Gelijke Behandeling (Dutch Equal Treatment Act) (n 23), art 2.3.
53. Algemene Wet Gelijke Behandeling (Dutch Equal Treatment Act) (n 23), art 2.3.
54. Tobias Nowak, 'Dutch Positive Action Measures in Higher Education in the Light of EU Law' (2022) 29 *Maastricht Journal of European and Comparative Law* 468.
55. Eindhoven University of Technology, 'TU/e resumes preferential policy for hiring female scientists' (19 April 2021) https://www.tue.nl/en/news/news-overview/tue-resumes-preferential-policy-for-hiring-female-scientists/ accessed 22 April 2025.
56. Weerts and others (n 2).
57. *See* Solon Barocas, Moritz Hardt and Arvind Narayanan, *Fairness and Machine Learning: Limitations and Opportunities* (MIT Press 2023) 340.
58. For example: Case C-407/98 *Katarina Abrahamsson and Leif Anderson v Elisabet Fogelqvist* ECLI:EU:C:2000:367 (CJEU, 6 July 2000); Case C-409/95 *Hellmut Marschall v Land Nordrhein-Westfalen* ECLI:EU:C:1997:533 (CJEU, 11 November 1997).
59. College voor de Rechten van de Mens (n 4) para 5.14.

Under the Dutch Equal Treatment Act, indirect discrimination is defined as follows.

> Indirect discrimination: if an apparently neutral provision, criterion or practice particularly affects persons of a particular religion, belief, political affiliation, race, sex, nationality, heterosexual or homosexual orientation or marital status compared to other persons.[60]

As noted, the Institute held that Breeze's matching algorithm discriminates indirectly against dark-skinned users.[61] The Institute followed Breeze's assumptions that Breeze's matching algorithm differentiated based on ethnicity.

However, it is not a given that Breeze does indeed put dark-skinned users at a disadvantage in the legal sense. Breeze based themselves on feedback from some users that the recommendations were overwhelmingly light-skinned. Statistical analysis would be necessary to establish that Breeze disadvantages dark-skinned users. We do not have access to information that enables us to do such an analysis. Even if Breeze's algorithm discriminates indirectly at first glance, this may be legal, namely if Breeze can rely on an objective justification.

The possibility for an objective justification (for prima facie indirect discrimination) is phrased as follows in the Dutch Equal Treatment Act:

> The prohibition of discrimination laid down in this law does not apply in respect of indirect discrimination if that distinction is objectively justified by a legitimate aim and the means of achieving that aim are appropriate and necessary.[62]

Hence, Breeze would have three criteria to check:

i. Does Breeze have a legitimate aim to use the algorithm?
ii. If so, is the algorithm an appropriate means to achieve that aim?
iii. If so, is the algorithm necessary to achieve that aim?

Tobler provides an excellent overview of the nuances of indirect discrimination in European law and case law; here we will summarize her findings and sketch an application to the Breeze case. [63] We discuss each element in turn.

(i) Does Breeze have a legitimate aim to use the algorithm?
Generally, case law shows that courts are willing to accept many aims as legitimate, and that questions (ii) and (iii) pose a more significant challenge. There are some limitations to the legitimate aims the CJEU has accepted. Particularly, in Szpital Kliniczny, the court found that a purely economic aim did not suffice.[64] Furthermore, in WABE and Muller, the CJEU found that there must be a genuine need for the aim provided.[65]

60. Algemene Wet Gelijke Behandeling (Dutch Equal Treatment Act) (n 23), art. 1(1)(c); The provision is based on Article 2(b), Council Directive 2000/43/EC 2009 22; See Council Directive 2000/78/EC of 27 November 2000 establishing a general framework for equal treatment in employment and occupation [2000] OJ L303/16, art 2(b); See generally on indirect discrimination, Christa Tobler, *Indirect Discrimination under Directives 2000/43/EC and 2000/78/EC: Towards a Harmonized European Approach?* (Intersentia 2005) 516.
61. College voor de Rechten van de Mens (n 4) para 5.11, 5.13.
62. Algemene Wet Gelijke Behandeling (Dutch Equal Treatment Act) (n 23), art 2(1).
63. Christa Tobler, *Indirect Discrimination under Directives 2000/43 and 2000/78* (European network of legal experts in gender equality and non-discrimination, Publications Office of the European Union 2022) 156.
64. Case C-16/19 *VL v Szpital Kliniczny im. dra J. Babińskiego Samodzielny Publiczny Zakład Opieki Zdrowotnej w Krakowie* ECLI:EU:C:2021:64 (Grand Chamber, 26 January 2021).
65. Joined Cases C-804/18 and C-341/19 *IX v WABE eV and MH Müller Handels GmbH v MJ* ECLI:EU:C:2021:594 (Grand Chamber, 15 July 2021).

The matching algorithm aims to provide users with adequate partners to go on a date with. This improves user experience, and is central to Breeze's business model, as Breeze gets paid per date. There is no clear legal precedent to point to in order to establish this as legitimate aim. Since Breeze's aim is not purely economic, and the algorithm is central to Breeze's business model, it seems plausible that running a dating app is a legitimate reason to use a matching algorithm.

(ii) Tobler finds that under the Court's case law, a measure taken in view of a legitimate aim is appropriate if it can indeed achieve the aim in question. In this case, Breeze would have to show that their algorithm improves the number of users finding a match. From a business perspective, it is likely that Breeze has investigated the algorithm's performance, and that Breeze could indeed demonstrate that their algorithm provides significantly more matches than randomly suggesting user profiles to each other. For now, let us assume that Breeze passes this hurdle.

(iii) The most contentious part of this assessment would then be whether the algorithm is 'necessary' for providing users with matches. This step of the assessment requires a proportionality test. Here, one must investigate whether the disadvantages caused to dark-skinned users are proportionate to the aim achieved by Breeze. For this test, one must consider the possibility of replacing the algorithm with a different way to achieve the aim of matching users, with fewer negative consequences for dark-skinned users. One could compare the Breeze algorithm to the algorithms of other dating apps, or any other economically feasible algorithm that is available to Breeze.[66] There is a lack of transparency from most dating apps, and a lack of scientific literature on matching algorithms for romantic purposes.[67] Therefore, the question of proportionality could focus, among other things, on Breeze demonstrating that they have investigated other options to the best of their abilities.

Following the above argumentation, it is feasible that Breeze could rely on objective justification for the prima facie indirect discrimination by Breeze's matching algorithm. This would be the case if a court agrees that matching users is a legitimate aim, and if a court agrees that the algorithm is appropriate and necessary. In sum, it is possible that Breeze (and its algorithm) never engaged in illegal indirect discrimination. But, as noted, the Institute did not examine the possibility of an objective justification, because the Institute focused on the questions asked by Breeze.

### 3.5 After the Breeze Case

Since the Breeze case, the Dutch Institute for Human Rights has dealt with several other cases in which it applied non-discrimination law to cases related to digital systems. One month after the Breeze decision, the institute decided in a case about the Free University in Amsterdam and online proctoring software called Proctorio. A dark-skinned student alleged that a university discriminated against her because the software did not properly recognise her face during online exams. However, the university proved that its software did not disadvantage or harm the student, at least no more than students with a light skin, by using the software. Hence, the university rebutted the presumption of indirect digital discrimination.[68]

66. Case C-83/14 *CHEZ Razpredelenie Bulgaria AD v Komisia za zashtita ot diskriminatsia* ECLI:EU:C:2015:480 (CJEU, 16 July 2015), para 127–130.

67. See for a US-focused discussion of discrimination in dating apps: Apryl Williams, *Not My Type: Automating Sexual Racism in Online Dating* (Stanford University Press 2024) 268; However, she does not discuss the details of the algorithms, as companies tend to keep the workings of their algorithms secret.

68. College voor de Rechten van de Mens, 'De Vrije Universiteit discrimineert een student niet door voor online tentamens de software van Proctorio te gebruiken. De Vrije Universiteit heeft de discriminatieklacht van de studente onvoldoende zorgvuldig behandeld.' ['The Free University did not discriminate against a student by using Proctorio software for online exams. The Free University did not handle the student's discrimination complaint with sufficient care.'] *See* Netherlands Institute for Human Rights, Decision 2023-111 (17 October 2023) https://oordelen.mensenrechten.nl/oordeel/2023-111 accessed 22 April 2025.

Also in the month after the Breeze decision, the Institute decided in three cases related to the Dutch childcare benefits scandal. In brief, the Institute decided that the Dutch tax office had discriminated indirectly against three people.[69] However, AI or digital systems hardly played a role in these cases.

In 2025, the Institute decided in a case about Facebook (Meta) and online advertising. [70] Roughly summarised, research showed that Facebook shows ads for stereotypically female jobs, such as receptionist and primary school teacher, mainly to women. Facebook showed ads for stereotypically male jobs, such as mechanic and electrician, mainly to men. The Institute decided that Facebook discriminated indirectly against women. Here, too, the Institute chose a broad interpretation of the concept 'service', claiming that offering a platform where people can pay to place their adverts constitutes a service.

## 4. Measuring Digital Discrimination

Suppose that a company wants to measure and mitigate discrimination in the digital system. In such a situation, it is crucial to have a conception of discrimination. A fairness intervention should not count as positive discrimination if it addresses previously existing discrimination in the system. Whether indirect discrimination is illegal depends partly on the question whether there a different system could be used that achieves the same goal with less harm. Both legal considerations rely on the *measurement* of discrimination. Current law does not give rules on how to measure discrimination in digital systems.

Computer scientists have developed many ways to define discrimination in numbers.[71] Computer scientists working on fair and non-discriminatory algorithms often speak of 'fairness metrics', which roughly translates 'ways to calculate a score which aims to capture the extent of discrimination in a system'. Frequently, number-based definitions of discrimination are in conflict: if one definition indicates that there is no discrimination, the other definition necessarily indicates that there is discrimination. Such conflicts can also happen if both number-based definitions of discrimination sound plausible in the abstract.[72]

Such conflicting definitions have resulted in several high-profile discussions on the correct way to measure discrimination. Famously, ProPublica reported on discrimination found in COMPAS, an algorithm to judge recidivism risk for defendants in the American court system.[73] Flores et al. used a different definition of fairness to conclude COMPASS is fair.[74] Initially, both approaches appear like valid attempts to measure discrimination; a more fine-grained approach is necessary to assess what the most appropriate definition of fairness is for a particular situation. It is likely that future cases of digital discrimination will spark debate on the correct way to measure discrimination. In this section, we outline the main definitions of

69. College voor de Rechten van de Mens, 'Drie oordelen in zaken van toeslagenouders: geen directe maar wel indirecte discriminatie door Belastingdienst/Toeslagen' ['Three judgements in cases of benefits parents: no direct but indirect discrimination by Tax Office/Benefits'] *See* Netherlands Institute for Human Rights, Decision 2023-101 (2023) https://oordelen.mensenrechten.nl/oordeel/2023-101 accessed 22 April 2025; Netherlands Institute for Human Rights, Decision 2023-102 (2023) https://oordelen.mensenrechten.nl/oordeel/2023-102 accessed 22 April 2025; Netherlands Institute for Human Rights, Decision 2023-103 (2023) https://oordelen.mensenrechten.nl/oordeel/2023-103accessed 22 April 2025.

70. College voor de Rechten van de Mens, 'Meta Platforms Ireland Ltd. maakt verboden onderscheid op grond van geslacht bij het tonen van advertenties voor vacatures aan gebruikers van Facebook in Nederland.' ['Meta Platforms Ireland Ltd. Discriminates indirectly and illegaly when showing job ads to Facebook users in the Netherlands.'] *See* Netherlands Institute for Human Rights, *Decision 2025-17* (18 February 2025) https://oordelen.mensenrechten.nl/oordeel/2025-17 accessed 22 April 2025.

71. Shira Mitchell, Eric Potash, Solon Barocas, Alexander D'Amour and Kristian Lum, 'Algorithmic Fairness: Choices, Assumptions, and Definitions' (2021) 8 *Annual Review of Statistics and Its Application* 141.

72. Alexandra Chouldechova, 'Fair Prediction with Disparate Impact: A Study of Bias in Recidivism Prediction Instruments' (2017) 5 *Big Data* 153; Jon Kleinberg, Sendhil Mullainathan and Manish Raghavan, 'Inherent Trade-Offs in the Fair Determination of Risk Scores' (2016) http://arxiv.org/abs/1609.05807 accessed 22 April 2025.

73. Julia Angwin, Jeff Larson, Surya Mattu and Lauren Kirchner, 'Machine Bias' (ProPublica, 23 May 2016) https://www.propublica.org/article/machine-bias-risk-assessments-in-criminal-sentencing accessed 22 April 2025.

74. Anthony W. Flores, Kristin Bechtel and Christopher T. Lowenkamp, 'False Positives, False Negatives, and False Analyses: A Rejoinder to "Machine Bias: There's Software Used Across the Country to Predict Future Criminals. And It's Biased Against Blacks ".' (2016) 80(2) *Federal Probation* 38.

digital discrimination. We show that it is often difficult to choose which way of defining and measuring discrimination is the most appropriate in a specific situation.

### 4.1 Institute's suggestion

We shall use the Institute's decision as a starting point: the Institute clarified that Breeze discriminates if it shows the profiles of dark-skinned users comparatively less frequently. Additionally, the Institute posits that users are allowed to have their preferences, but that it is not permissible for Breeze to *amplify* those preferences.[75] It is unclear from this statement whether it is permissible for a dating app to *reproduce* user preferences, if it does not amplify them.

To interpret the suggestion from the Breeze decision, we rely on literature from the field of fair machine learning.[76] A dating app uses a *recommender system*. Just as the web has become integral to our interaction with the world, recommender systems have become a central infrastructure in our interaction with the web. Spotify recommends songs, Amazon recommends objects to be sold, Google recommends web pages, and LinkedIn recommends people and jobs. This makes the study of recommender systems, Information Retrieval, a wide and varied field: HR requires a different understanding of quality and fairness than single recommendation, which is in turn quite different from finding the right scientific document matching a search term. The core task of recommender systems is to provide users with the information they desire. Ekstrand et al. phrase this as: 'Given a repository of items and a user information need, present items to help the user satisfy that need.'[77]

There is a rich literature investigating *fairness* in recommender systems.[78] Although recommender systems are frequently used in contexts where benefits are being distributed, the recommender system itself does not allocate those goods. A recommender system in a marketplace only shows users the goods on the marketplace. As such, the analysis of recommender systems focuses on *exposure*:[79] exposure quantifies how frequently a user sees a particular item. This is in line with the Institute's decision.

### 4.2 Demographic Parity

We shall use the Institute's decision as a starting point: the Institute clarified that Breeze discriminates if it shows the profiles of the dark-skinned user comparatively less frequently. Although this statement appears conceptually clear, this comparison is difficult to operationalize. A plausible first interpretation is the fairness metric called Demographic Parity, or Equal Treatment. Demographic Parity holds that different groups (demographics) should, on average, obtain equal outcomes.

In the context of the Breeze case, Demographic Parity holds that the algorithm should show the average profile of a dark-skinned user and the average profile of a light-skinned user equally often. Demographic Parity is easy to understand and implement, and seems fitting in the context of romance. Dark-skinned and

75. College voor de Rechten van de Mens (n 4) para 5.13.

76. Many scholars working on fair machine learning meet every year at the FAccT Conference; FAccT Conference, 'FAccT: Conference on Fairness, Accountability, and Transparency' https://facctconference.org accessed 22 April 2025; Papers on the questions mentioned in the main text include: Reuben Binns, 'Fairness in Machine Learning: Lessons from Political Philosophy' in Sorelle A Friedler and Christo Wilson (eds), *Proceedings of the Conference on Fairness, Accountability, and Transparency* (Proceedings of Machine Learning Research, vol 81, PMLR 2018) 149, 149–159; Pak-Hang Wong, 'Democratizing Algorithmic Fairness' [2020] 33 *Philosophy & Technology* 225; Pratik Gajane and Mykola Pechenizkiy, 'On Formalizing Fairness in Prediction with Machine Learning' (2017) arXiv preprint arXiv:1710.03184 https://arxiv.org/abs/1710.03184 accessed 22 April 2025.

77. Ekstrand and others (n 41).

78. Deldjoo and others (n 41); Yifan Wang, Ruobing Xie, Xiang Wang, Leyu Lin, Tat-Seng Chua and Min-Yen Kan, 'A Survey on the Fairness of Recommender Systems' (2023) 41 *ACM Transactions on Information Systems* 52:1.

79. Meike Zehlike and Carlos Castillo, 'Reducing Disparate Exposure in Ranking: A Learning To Rank Approach' in *Proceedings of The Web Conference 2020* (Association for Computing Machinery 2020) 2849–2855 https://dl.acm.org/doi/10.1145/3366424.3380048 accessed 24 May 2024; Fernando Diaz, Bhaskar Mitra, Michael D. Ekstrand, Asia J. Biega and Ben Carterette, 'Evaluating Stochastic Rankings with Expected Exposure' in *Proceedings of the 29th ACM International Conference on Information and Knowledge Management* (Association for Computing Machinery 2020) 137–146 https://dl.acm.org/doi/10.1145/3340531.3411962 accessed 24 May 2024; Ashudeep Singh and Thorsten Joachims, 'Fairness of Exposure in Rankings' in *Proceedings of the 24th ACM SIGKDD International Conference on Knowledge Discovery & Data Mining* (Association for Computing Machinery 2018) 2219–2228 https://dl.acm.org/doi/10.1145/3219819.3220088 accessed 24 May 2024.

light-skinned users have equal rights to exposure on a dating app. To enforce Demographic Parity, a dating app needs to ensure that their matching algorithm gives equal exposure to dark-skinned and light-skinned users on average.[80] Despite its convenience, fair machine learning scholars have identified several problems with Demographic Parity.[81]

The first problem with Demographic Parity is that a system that enforces Demographic Parity might perform worse for the average user than a system that does not.[82] To an extent, it holds more broadly that if you alter a system to adhere to a fairness metric, the system performs less well for the average user.[83] Demographic Parity causes a relatively bad drop in performance due to its crudeness.

If a dating app stops their algorithm from discriminating, users might have to wait longer to match with other users. This presents a difficult challenge for machine learning practitioners; they have to weigh the interests of a disadvantaged minority against the interests of the population as a whole. If users have to wait too long to find a match, they might stop using the app altogether, which can hurt the business.

Discrimination law provides a broad answer to this question, namely that a difference in treatment of different groups is permissible if there is no solution that produces less harm, but still achieves the aim of the system. Again, this answer might be too general to quantify. A marginal decrease of discrimination probably is not proportionate to a much worse experience for users overall. On the other hand, erasing discrimination in the system entirely does justify a marginal decrease in overall performance. Somewhere between those two extremes, there is a grey area, which is too context-dependent to quantify generally. Scholars have argued against quantifying this tradeoff, even if it were possible.[84] There does not appear to be a simple or general way to balance the interests of a company, the disadvantaged minority, and the population in its entirety.

The second problem with Demographic Parity is that there may be legitimate reasons why groups receive different outcomes.[85] Demographic Parity requires that each group receives, on average, the same

80. Faisal Kamiran, Indrė Žliobaitė and Toon Calders, 'Quantifying Explainable Discrimination and Removing Illegal Discrimination in Automated Decision Making' (2013) 35 *Knowledge and Information Systems* 613; Muhammad Bilal Zafar, Isabel Valera, Manuel Gomez Rodriguez, Krishna P Gummadi and Adrian Weller, 'From Parity to Preference-Based Notions of Fairness in Classification' in *Advances in Neural Information Processing Systems 30 (NeurIPS 2017)* (Curran Associates Inc 2017) 229–239 https://proceedings.neurips.cc/paper_files/paper/2017/hash/82161242827b703e6acf9c726942a1e4-Abstract.html accessed 9 May 2025; Michael Feldman, Sorelle A Friedler, John Moeller, Carlos Scheidegger and Suresh Venkatasubramanian, 'Certifying and Removing Disparate Impact' in *Proceedings of the 21st ACM SIGKDD International Conference on Knowledge Discovery and Data Mining* (Association for Computing Machinery 2015) 259–268 https://dl.acm.org/doi/10.1145/2783258.2783311 accessed 9 May 2025.

81. James R. Foulds and Shimei Pan, 'Are Parity-Based Notions of AI Fairness Desirable?' (2020) 43(4) *IEEE Data Engineering Bulletin* 51–73 https://www.semanticscholar.org/paper/Are-Parity-Based-Notions-of-AI-Fairness-Desirable-Foulds-Pan/b811870a7aa2b806bb51cbac2f149bd27566a474 accessed 27 March 2024; Elizabeth Anne Watkins and Jiahao Chen, 'The Four-Fifths Rule Is Not Disparate Impact: A Woeful Tale of Epistemic Trespassing in Algorithmic Fairness' in *Proceedings of the 2024 ACM Conference on Fairness, Accountability, and Transparency (FAccT '24)* (Association for Computing Machinery 2024) 764–775 https://dl.acm.org/doi/10.1145/3630106.3658938 accessed 9 May 2025.

82. Moritz Hardt, Eric Price and Nathan Srebro, 'Equality of Opportunity in Supervised Learning' in *Advances in Neural Information Processing Systems 29 (NeurIPS 2016)* (Curran Associates Inc 2016) 3315–3323; Andrew Bell, Lucius Bynum, Nazarii Drushchak, Tetiana Herasymova, Lucas Rosenblatt and Julia Stoyanovich, 'The Possibility of Fairness: Revisiting the Impossibility Theorem in Practice' in *Proceedings of the 2023 ACM Conference on Fairness, Accountability, and Transparency (FAccT '23)* (Association for Computing Machinery 2023) 400–412 https://doi.org/10.1145/3593013.3594007 accessed 1 May 2025.

83. Bell and others (n 82); Dimitris Bertsimas, Vivek F Farias and Nikolaos Trichakis, 'On the Efficiency-Fairness Trade-Off' (2012) 58 *Management Science* 2234–2250; Sepehr Dehdashtian, Bashir Sadeghi and Vishnu Naresh Boddeti, 'Utility-Fairness Trade-Offs and How to Find Them' in *Proceedings of the IEEE/CVF Conference on Computer Vision and Pattern Recognition (CVPR 2024)* (IEEE 2024) 12038–12046 https://openaccess.thecvf.com/content/CVPR2024/html/Dehdashtian_Utility-Fairness_Trade-Offs_and_How_to_Find_Them_CVPR_2024_paper.html accessed 9 May 2025.

84. Hilde Weerts, Raphaële Xenidis, Fabien Tarissan, Henrik Palmer Olsen and Mykola Pechenizkiy, 'Algorithmic Unfairness through the Lens of EU Non-Discrimination Law: Or Why the Law Is Not a Decision Tree' in *Proceedings of the 2023 ACM Conference on Fairness, Accountability, and Transparency (FAccT '23)* (Association for Computing Machinery 2023) 805–816 https://dl.acm.org/doi/10.1145/3593013.3594044 accessed 7 May 2025; Andrew D. Selbst, Danah Boyd, Sorelle A. Friedler, Suresh Venkatasubramanian and Janet Vertesi, 'Fairness and Abstraction in Sociotechnical Systems' in *Proceedings of the 2019 Conference on Fairness, Accountability, and Transparency (FAT '19)* (Association for Computing Machinery 2019) 59–68 https://dl.acm.org/doi/10.1145/3287560.3287598 accessed 7 May 2025.

85. Foulds and Pan (n 81); Sam Corbett-Davies, Johann D. Gaebler, Hamed Nilforoshan, Ravi Shroff and Sharad Goel, 'The Measure and Mismeasure of Fairness' (2024) 24 *Journal of Machine Learning Research* 312:14730.

outcomes. This can lead to difficulties: we can imagine a scenario in the realm of dating. Minority A consists of very efficient daters – after one date, they typically find their match, and stop using the app. Minority B, by contrast, wants to explore, and dates with several people before quitting the app. Once someone has stopped using the app, the algorithm will stop showing their profile. The algorithm thus shows Minority A less frequently than Minority B, as Minority A is quicker to stop using the app. The algorithm's behaviour would violate Demographic Parity, which requires the algorithm to show both minorities equally frequently. Yet, it is unclear that either minority is harmed.

### 4.3 Conditional Demographic Parity

Conditional Demographic Parity holds that demographics should receive – on average – equal outcomes, if they are equal in some respect. [86] In the previous example, one could argue that users are only harmed if they are shown less than their activity level would suggest. Conditional Demographic Parity tries to capture such intuitions. In the previous example, Demographic Parity would require the system to continue showing inactive profiles from Minority A, or to stop showing profiles from Minority B even though they want to continue dating. Conditional Demographic Parity allows for a more specific comparison: we compare those in Minority A *that leave the app after one date* to those in Minority B *that leave the app after one date*. We can interpret this as Demographic Parity, *conditional* on the *attribute* activity level.

It can be difficult to apply Conditional Demographic Parity to more than one attribute at a time.[87] To illustrate: it might be possible to compare people of different ethnicities that have the same income, *or* equal height, but there might not be enough people with equal height *and* income to meaningfully compare. This limits Conditional Demographic Parity as a measuring tool.

Since Conditional Demographic Parity only allows for looking at a limited number of attributes, much depends on the choice of attributes. In the previous example, 'activity level' appeared to be a reasonable attribute: if users do not want to use an app, certainly, the app shouldn't need to show their profiles. However, users might become more or less active depending on their level of success. As such, if an app shows the profiles of one demographic less frequently, that demographic might well become less active. Even this attribute that appears neutral, might still be a direct result of discrimination.

### 4.4 Equity

Another group of fairness metrics looks at merit. For example, in mortgage finance, it is common and not too controversial that banks differentiate on income and savings to establish the size of mortgage one can repay. As different demographics can have different financial circumstances, this differentiation can result in indirect discrimination. It is not clear that a bank causes less harm for a disadvantaged group if they started giving out larger mortgages. These larger mortgages could lead to many defaults, causing serious harm to the disadvantaged group, and society in a broader sense. In this case, it might be sensible for banks to define some sense of *merit* – here, how much money a borrower could repay – and to lend money according to this merit.

Equity is one such fairness metric[88]. Equity holds that demographics should receive outcomes proportionate to their merit. Recommender systems frequently hold some notion of equity, particularly with popularity as merit. If book A is more popular than book B, it does not seem unfair if book A is promoted more frequently than book B. Equity becomes a fairness metric when we compare different demographics, rather than different books. Again, there are many ways to exactly calculate equity. One could require that demographics

86. Sandra Wachter, Brent Mittelstadt and Chris Russell, 'Why Fairness Cannot Be Automated: Bridging the Gap between EU Non-Discrimination Law and AI' (2021) 41 *Computer Law & Security Review* 105567.

87. Michael Kearns, Seth Neel, Aaron Roth and Zhiwei Steven Wu, 'Preventing Fairness Gerrymandering: Auditing and Learning for Subgroup Fairness' (2018) 80 *Proceedings of the 35th International Conference on Machine Learning* 2564–2572 https://proceedings.mlr.press/v80/kearns18a.html accessed 24 May 2024.

88. Asia J. Biega, Krishna P. Gummadi and Gerhard Weikum, 'Equity of Attention: Amortizing Individual Fairness in Rankings' in *Proceedings of the 41st International ACM SIGIR Conference on Research & Development in Information Retrieval* (ACM 2018) 405–414 https://dl.acm.org/doi/10.1145/3209978.3210063 accessed 24 May 2024.

receive exposure proportional to their popularity[89], but more complex calculations are possible.[90] Equity appears less fitting in the dating market, as applying equity in the dating market implies that different demographics have different merits.

### 4.5 Notions of Fairness

The notions of fairness we have introduced so far do not align: indeed, various impossibility theorems detail how a system cannot fulfil two fairness notions at the same time, outside of exceptional circumstances.[91] This is no great worry for fairness measurement, as it mostly means that these notions of fairness capture different concepts. One cannot fulfil both Demographic Parity and Equity, as the groups cannot get *both* equal exposure *and* exposure proportional to merit.

'*Causal Fairness*' aims to establish all relevant causes and effects from ethnicity to the measured outcome.[92] This would allow an expert to indicate some cause-effect relations as acceptable (e.g. it is okay if a light-skinned person's profile gets shown less if the person is inactive), and some as unacceptable (e.g. it is not okay to show a dark-skinned person's profile less if your population is racist). The unacceptable cause-effect relations could then be removed from the model, resulting in a model in which each step is acceptable. Causal fairness is conceptually promising, but requires a detailed model of cause and effect to function. The outcome of the investigation depends greatly on the choice of causal model. Many situations allow for different causal models. Causal fairness can facilitate the discussion on how to measure discrimination in models, more than that causal fairness measures fairness.

In sum, there are many ways to define fairness, and even more ways to measure fairness. The Institutes claim that it is discriminatory if Breeze shows the profiles of dark-skinned users less frequently than those of light-skinned users can be quantified in many ways. Demographic Parity suggests we compare the average dark-skinned user to the average light-skinned user. Conditional Demographic Parity allows us to compare these populations, mediated by some attribute. Equity measures the average dark-skinned user receives attention proportionate to their popularity. Each of these statements is a brutish summarization of the greater variety underlying each of these terms. Different measurements are more suited to different contexts, but for any context, there can be many plausible possibilities to measure discrimination.

## 5. Mitigating Discrimination

The Institute concluded that regardless of whether the discrimination was illegal, Breeze is allowed (and required) to mitigate the discrimination in their system. In this section, we will investigate how to mitigate discrimination in a dating app.

### 5.1 Accessing Ethnicity Data

It seems natural to rely on the Fair Machine Learning literature, which has a broad purpose to limit discrimination in algorithmic systems. The field of Fair Machine Learning stems from Computer Science and is quite technical. Research of the Fair Machine Learning community is important, as discrimination by algorithmic systems is a serious societal problem. Unfortunately, the practical applicability of many fairness interventions is limited.

89. Singh and Joachims (n 79); Biega, Gummadi and Weikum (n 88).

90. Piotr Sapiezynski, Wesley Zeng, Ronald E Robertson, Alan Mislove and Christo Wilson, 'Quantifying the Impact of User Attention on Fair Group Representation in Ranked Lists' in *Companion Proceedings of The 2019 World Wide Web Conference* (ACM 2019) 923–928 https://dl.acm.org/doi/10.1145/3308560.3317595 accessed 9 May 2025; Amifa Raj and Michael D Ekstrand, 'Comparing Fair Ranking Metrics' (2020) arXiv preprint arXiv:2009.01311 https://arxiv.org/abs/2009.01311 accessed 9 May 2025.

91. Chouldechova (n 72); Kleinberg, Mullainathan and Raghavan (n 72); Fabian Beigang, 'Yet Another Impossibility Theorem in Algorithmic Fairness' (2023) 33 *Minds and Machines* 715.

92. Matt J. Kusner, Joshua R. Loftus, Chris Russell and Ricardo Silva, 'Counterfactual Fairness' in *Advances in Neural Information Processing Systems 30 (NeurIPS 2017)* (Curran Associates Inc 2017) 4066–4076 https://papers.nips.cc/paper/2017/hash/a486cd07e4ac3d270571622f4f316ec5-Abstract.html accessed 24 May 2024.

For instance, many fairness interventions require access to ethnicity (or whichever attribute is being discriminated on) to measure discrimination. However, the GDPR bans, in principle, using ethnicity data. Article 9.1 of the GDPR holds that 'Processing of personal data revealing racial or ethnic origin [...] shall be prohibited.' As a result, Breeze cannot use the ethnicity of users in its attempts to mitigate discrimination. This places Breeze in a difficult position; to prevent discrimination, they would need to record ethnicity to measure discrimination, but recording ethnicity breaches the GDPR.[93]

There are some possibilities to circumvent this prohibition. Article 9.2a of the GDPR holds that the ban on using ethnicity data 'shall not apply if the data subject has given explicit consent to the processing of those personal data for one or more specified purposes [...]'. At first glance, this looks like a plausible way for Breeze to record their users' ethnicities. Unfortunately, this is probably insufficient. Some users might be unwilling to volunteer their ethnicity, for example for privacy reasons. This means that Breeze would get data from only a sample of the population, without means to test whether this sample is representative of the larger population. If the sample is not representative of the population, conclusions drawn from analysis of the sample might not hold for the population at large, which makes the analysis lose power. Unless self-reported ethnicity data can provide a reliably representative sample, self-reported ethnicity data cannot provide a complete answer to digital discrimination. The GDPR includes more exceptions to the ban on using ethnicity data, but none of the exceptions is suitable for de-biasing algorithms.[94]

Article 10.5 of the AI Act states that developers of 'high-risk' AI systems may process their users' ethnicities for de-biasing their systems, under certain strict conditions.[95] However, the AI Act does not include dating apps in the list of high-risk AI.[96] Hence, that de-biasing provision does not apply, and cannot help providers of dating apps.

Although this prohibition on the use of ethnicity data disallows the mitigation of discrimination acting on a direct measurement, this is not necessarily a bad thing. Selbst et al. identify the Formalism Trap: 'Failure to account for the full meaning of social concepts such as fairness, which can be procedural, contextual, and contestable, and cannot be resolved through mathematical formalisms.'[97] Companies might only ensure their systems satisfy a fixed measurement of discrimination rather than mitigating the underlying harm.[98] As Goodhart's Law goes: 'When a measure becomes a target, it ceases to be a good measure.'[99]

### 5.2 Mitigating Discrimination without Ethnicity Data

One of the main pillars of Breeze's argument was that popular users get recommended disproportionately much (popularity bias). Rather than trying to measure discrimination and directly influencing the outcome of that measurement, Breeze could aim to prevent this popularity bias. Eliminating popularity bias in the matching algorithm would then also eliminate the part of discrimination caused by that popularity bias.

93. Marvin Van Bekkum and Frederik Zuiderveen Borgesius, 'Using Sensitive Data to Prevent Discrimination by Artificial Intelligence: Does the GDPR Need a New Exception?' (2023) 48 *Computer Law & Security Review* 105770.

94. Van Bekkum and Borgesius (n 93).

95. Regulation (EU) 2024/1689 of the European Parliament and of the Council of 13 June 2024 laying down harmonised rules on artificial intelligence and amending Regulations (EC) No 300/2008, (EU) No 167/2013, (EU) No 168/2013, (EU) 2018/858, (EU) 2018/1139 and (EU) 2019/2144 and Directives 2014/90/EU, (EU) 2016/797 and (EU) 2020/1828 (Artificial Intelligence Act) [2024] OJ L 2024/1689.

96. See for an in-depth analysis of Article 10.5 of the AI Act: Marvin van Bekkum, 'Using sensitive data to de-bias AI systems: Article 10(5) of the EU AI Act' [2025] 56 *Computer Law & Security Review* 106115.

97. Selbst and others (n 84).

98. Tim De Jonge and Djoerd Hiemstra, 'UNFair: Search Engine Manipulation, Undetectable by Amortized Inequity' in *Proceedings of the 2023 ACM Conference on Fairness, Accountability, and Transparency* (Association for Computing Machinery 2023) https://dl.acm.org/doi/10.1145/3593013.3594046 accessed 27 March 2024.

99. Marilyn Strathern, '"Improving Ratings": Audit in the British University System' (1997) 5 *European Review* 305.

Popularity bias is common in many types of recommender systems, as popularity bias[100], the Matthew effect[101], or rich-get-richer dynamics[102]. Consequently, there is also a broad literature of papers addressing popularity bias in recommender systems. By and large, such research focuses on one-sided recommendation: recommending movies[103], or news[104], to users. Dating has different considerations, as dating apps recommend users to other users. Future research will have to demonstrate whether these solutions can also address popularity bias in dating apps, and whether this does indeed mitigate discrimination.

There is also a modest literature addressing popularity bias in dating. Xia et al. take an economic approach, and suggest an algorithm that prioritizes market clearance.[105] Their algorithm prioritizes showing those users for whom the algorithm estimates a low matching probability and so counters popularity bias. Pizzato and Silvestrini suggested making all recommendations mutual.[106] Then, if we force the algorithm to make a certain amount of recommendations to each active user, it is also guaranteed that the algorithm shows each profile a minimum amount of times. While this might not entirely eliminate popularity bias, it does limit the possible variation in how frequently a profile is shown. Further research will have to demonstrate the effectiveness of this solution for less popular users.

Additionally, it is possible for Breeze to implement non-technical solutions.[107] As an example, Breeze identified that their user base mostly consists of light-skinned people. Breeze could address this through dialogue with dark-skinned people, to find if there are barriers keeping them from entry. In a previous version, Breeze only provided dates in bars, where alcohol is typically consumed. This could present a barrier of entry for some Muslim users, who do not consume alcohol, and who might not want to be seen in places so strongly associated with alcohol. Breeze has since amended this policy, and now Breeze also allows for dates where the users themselves choose where to go.[108]

Breeze could also alter their communication strategy to cater more towards dark-skinned users. Hutson et al. identifies community guidelines, behaviour agreements, or shareable media articles as several ways in which dating platforms could address discrimination through communication. [109] Additionally, if one of the causes of discrimination in this case was the fact that Breeze's user base was predominantly light-skinned, Breeze could cater their advertisement strategy more towards dark-skinned people.

100. Himan Abdollahpouri, Robin Burke and Bamshad Mobasher, 'Controlling Popularity Bias in Learning-to-Rank Recommendation' in *Proceedings of the Eleventh ACM Conference on Recommender Systems* (Association for Computing Machinery 2017) https://dl.acm.org/doi/10.1145/3109859.3109912 accessed 27 March 2024; Himan Abdollahpouri, Robin Burke and Bamshad Mobasher, 'Managing Popularity Bias in Recommender Systems with Personalized Re-Ranking' (2019) arXiv preprint arXiv:1901.07555 https://arxiv.org/abs/1901.07555 accessed 27 March 2024.
101. Hao Wang, Zonghu Wang and Weishi Zhang, 'Quantitative Analysis of Matthew Effect and Sparsity Problem of Recommender Systems' in *2018 IEEE 3rd International Conference on Cloud Computing and Big Data Analysis (ICCCBDA)* (IEEE 2018) 78–82 https://doi.org/10.1109/ICCCBDA.2018.8386490 accessed 27 March 2024.
102. Fabrizio Germano, Vicenç Gómez and Gaël Le Mens, 'The Few-Get-Richer: A Surprising Consequence of Popularity-Based Rankings?' in *Proceedings of the 2019 World Wide Web Conference* (ACM 2019) 2743–2749 https://doi.org/10.1145/3308558.3313693 accessed 27 March 2024; Marco Morik, Ashudeep Singh, Jessica Hong and Thorsten Joachims, 'Controlling Fairness and Bias in Dynamic Learning-to-Rank' in *Proceedings of the 43rd International ACM SIGIR Conference on Research and Development in Information Retrieval* (ACM 2020) 429–438 https://doi.org/10.1145/3397271.3401100 accessed 27 March 2024.
103. Abdollahpouri, Burke and Mobasher, 'Controlling Popularity Bias in Learning-to-Rank Recommendation' (n 100); Abdollahpouri, Burke and Mobasher, 'Managing Popularity Bias in Recommender Systems with Personalized Re-Ranking' (n 100).
104. Morik and others (n 102).
105. Bin Xia, Junjie Yin, Jian Xu and Yun Li, 'WE-Rec: A Fairness-Aware Reciprocal Recommendation Based on Walrasian Equilibrium' (2019) 182 *Knowledge-Based Systems* 104857 https://doi.org/10.1016/j.knosys.2019.07.028 accessed 27 March 2024.
106. Luiz Augusto Pizzato and Cameron Silvestrini, 'Stochastic Matching and Collaborative Filtering to Recommend People to People' in *Proceedings of the Fifth ACM Conference on Recommender Systems* (ACM 2011) 341–344.
107. Laurens Naudts and Anton Vedder, 'Fairness and Artificial Intelligence' in Nathalie A Smuha (ed), *The Cambridge Handbook of the Law, Ethics and Policy of Artificial Intelligence* (Cambridge University Press 2025) 79–100.
108. *See* 'FAQ' (Breeze Social) https://breeze.social/nl/faq accessed 8 May 2025.
109. Hutson and others (n 12).

Scholars have noted that, in general, technical interventions with fairness metrics are often insufficient or inappropriate to make a situation fair. In many cases, solutions must be sought outside fairness metrics.[110]

## 6. Conclusion

In this paper, we provided an overview of a decision in which the Netherlands Institute for Human Rights said that the company Breeze was allowed and required to fix the supposed discrimination in its app. The decision illustrates that general discrimination law can be applied to digital discrimination cases. The decision also illustrates that, in some circumstances, taking action to mitigate discrimination by a digital system does not count as positive discrimination. The lack of statistical information on the Breeze case makes it difficult to evaluate whether there was indirect discrimination, and if so, whether it was illegal. There are many different numerical definitions of discrimination, which do not necessarily agree. Even with statistics available, evaluating the legality of discrimination would be difficult. Moreover, in principle the GDPR prevents using ethnicity data, preventing the measurement of discrimination. Still, Breeze could address the discrimination in their system; either Breeze could address the popularity bias in their system, or Breeze could rely on non-technical interventions to increase diversity in their user population.

110. *See* Solon Barocas, Moritz Hardt and Arvind Narayanan, *Fairness and Machine Learning: Limitations and Opportunities* (MIT Press 2023) 238–247.

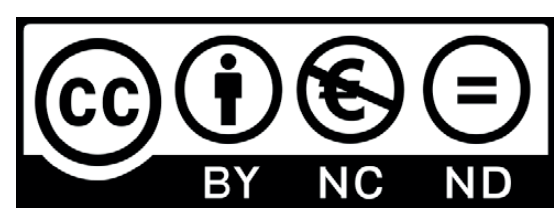